\def\year{2020}\relax
\documentclass[letterpaper]{article} 
\usepackage{aaai20}  
\usepackage{times}  
\usepackage{helvet} 
\usepackage{courier}  
\usepackage[hyphens]{url}  
\usepackage{graphicx} 
\usepackage{graphicx}  
\usepackage{epsfig}
\usepackage{graphicx}
\usepackage{amsmath}
\usepackage{amssymb}
\usepackage{bbm}
\usepackage{bm}
\usepackage{color}
\usepackage{multirow}
\usepackage{array}
\usepackage{booktabs}
\usepackage{algorithm}
\usepackage{algpseudocode}
\usepackage{amsmath}
\usepackage{diagbox}
\newcommand{\Tref}[1]{Table~\ref{#1}}

\newcommand{\Fref}[1]{Figure~\ref{#1}}

\newcommand{\etal}{\textit{et al.}}
\title{Model Watermarking for Image Processing Networks}

\author{ \Large \textbf{Jie Zhang$^\dagger$,\textsuperscript{\rm 1} Dongdong Chen$^\dagger$\thanks{Corresponding author, $\dagger$ Equal contribution.},\textsuperscript{\rm 2} Jing Liao,\textsuperscript{\rm 3} Han Fang,\textsuperscript{\rm 1}} \\\Large \textbf{Weiming Zhang,\textsuperscript{\rm 1} Wenbo Zhou,\textsuperscript{\rm 1} Hao Cui,\textsuperscript{\rm 1} Nenghai Yu\textsuperscript{\rm 1} }\\ 
\textsuperscript{\rm 1}University of Science and Technology of China  \ \textsuperscript{\rm 2}Microsoft Cloud AI  \   \textsuperscript{\rm 3}City University of Hong Kong\\
\textsuperscript{\rm 1}\{zjzac@mail., fanghan@mail., zhangwm@, welbeckz@mail., cvhc@mail., ynh@\}ustc.edu.cn\\
\textsuperscript{\rm 2}cddlyf@gmail.com  \ \textsuperscript{\rm 3}jingliao@cityu.edu.hk\\
}
  
\begin{document}

\maketitle

\begin{abstract}
Deep learning has achieved tremendous success in numerous industrial applications. As training a good model often needs massive high-quality data and computation resources, the learned models often have significant business values. However, these valuable deep models are exposed to a huge risk of infringements. For example, if the attacker has the full information of one target model including the network structure and weights, the model can be easily finetuned on new datasets. Even if the attacker can only access the output of the target model, he/she can still train another similar surrogate model by generating a large scale of input-output training pairs. How to protect the intellectual property of deep models is a very important but seriously under-researched problem. There are a few recent attempts at classification network protection only.  

In this paper, we propose the first model watermarking framework for protecting image processing models. To achieve this goal, we leverage the spatial invisible watermarking mechanism. Specifically, given a black-box target model, a unified and invisible watermark is hidden into its outputs, which can be regarded as a special task-agnostic barrier. In this way, when the attacker trains one surrogate model by using the input-output pairs of the target model, the hidden watermark will be learned and extracted afterward. To enable watermarks from binary bits to high-resolution images, both traditional and deep spatial invisible watermarking mechanism are considered. Experiments demonstrate the robustness of the proposed watermarking mechanism, which can resist surrogate models learned with different network structures and objective functions. Besides deep models, the proposed method is also easy to be extended to protect data and traditional image processing algorithms.

\end{abstract}

\section{Introduction}
In recent years, deep learning has revolutionized a wide variety of tasks such as image recognition \cite{krizhevsky2012imagenet,he2016deep}, medical image processing \cite{hong2017encase,hong2019combining,zhang2017rebuild,hong2017event2vec}, speech recognition \cite{graves2013speech,zhang2017towards} and natural language processing \cite{vaswani2017attention}, and significantly outperforms traditional state-of-the-art methods. To fully utilize the strong learning capability of these deep models and avoid overfitting, a large scale of high-quality labeled data and massive computation resources are often required. Since both human annotation and computation resources are expensive, these learned models are of great business value and need to be protected. But compared to traditional image watermarking techniques, protecting the intellectual property (IP) of deep models is much more challenging. Because of the exponential search space of network structures and weights, numerous structure and weight combinations exist for one specific task. In other words, we can achieve similar or better performance even if we slightly change the structure or weights of the target model.

In the white-box case, where the full information including the detailed network structure and weights of the target model is known, one typical and effective attacking way would be fine-tuning or pruning based on the target model on new task-specific datasets. Even in the black-box case, where only the output of the target model can be accessed, we can still steal the intellectual property of the target model by using another surrogate model to imitate its behavior. Specifically, we can first generate a large scale of input-output training pairs based on the target model then directly train the surrogate model in a supervised manner by regarding the outputs of the target model as ground-truth labels.

Very recently, some research works \cite{uchida2017embedding,adi2018turning,zhang2018protecting,merrer2017adversarial} start paying attention to the IP protection problem for deep neural networks. They often either add a parameter regularizer to the loss function or use the predictions of a special set of indicator images as the watermarks. However, deep watermarking is still a seriously under-researched field and all existing methods only consider the classification task. And in real scenarios, labeling the training data for image processing tasks, is much more complex and expensive than classification tasks, because their ground-truth labels should be pixel-wisely precise. Examples include removing all the ribs in Chest X-ray images and the rain streaks in real rainy images. Therefore protecting such image processing models is more valuable.

Motivated by this, this paper considers the deep watermarking problem for image processing models for the first time. And because the original raw model does not need to be provided in most application scenarios, they can be easily encrypted with traditional algorithms to resist the white-box attack (i.e., fine-tuned or pruned). So we mainly consider the second black-box attack case where only outputs of the target model can be obtained and attackers use surrogate models to imitate it. To resist such attacks, the designed watermarking mechanism should guarantee the watermarks can still be extracted from outputs of learned surrogate models.

\begin{figure}
    \centering
    \includegraphics[width=0.95\linewidth]{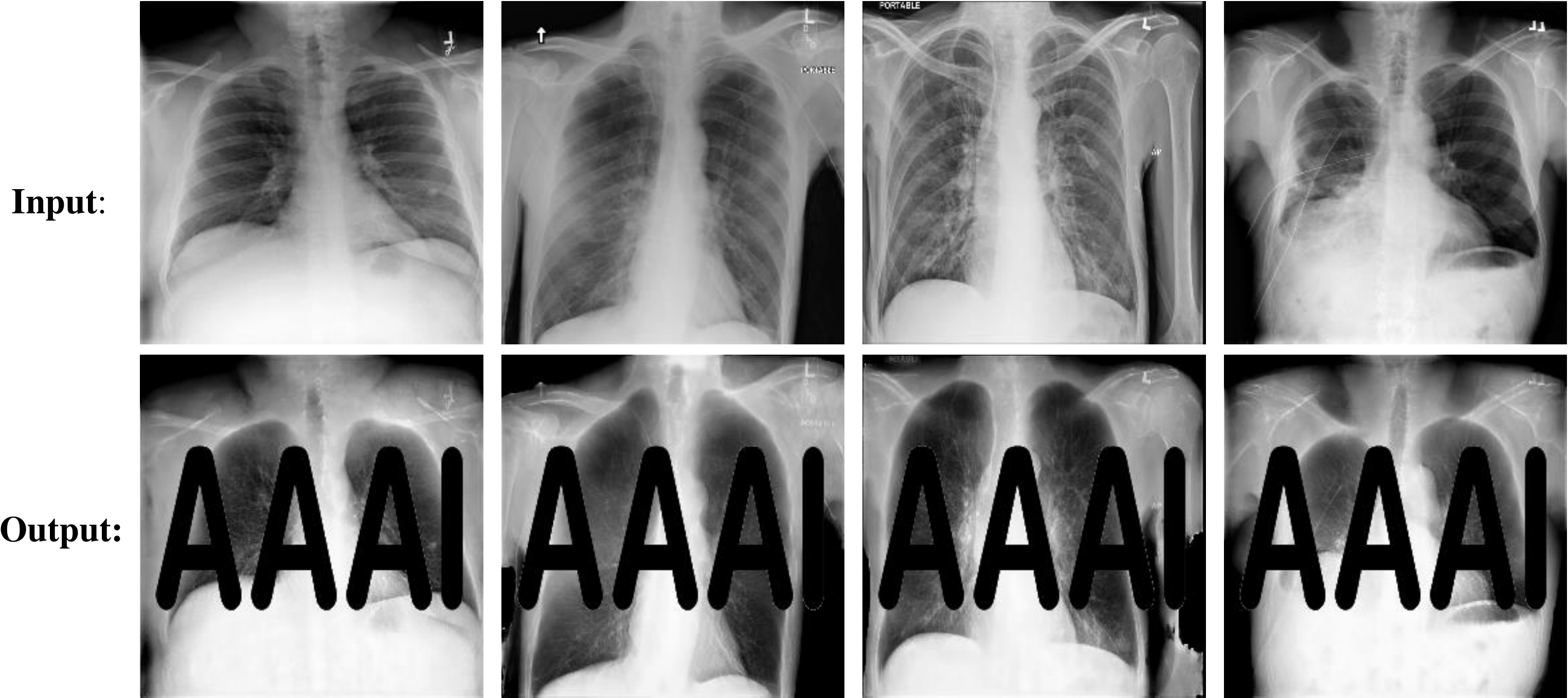}
    \caption{The simplest watermarking mechanism by adding unified visible watermarks onto the target output images, which will sacrifice the visual quality and usability.}
    \label{fig:simple_vis_wm}
\end{figure}

Before diving into the model watermarking for image processing networks, we first discuss the simplest spatial visible watermarking mechanism shown in \Fref{fig:simple_vis_wm}. Suppose that we have a lot of input-output training pairs and we manually add a unified visible watermark template to all the outputs. Intuitively, if a surrogate model is trained on such pairs with the simple L2 loss, the learned model will learn this visible watermark into its output to get lower loss. That is to say, given one target model, if we forcibly add one unified visible watermark into all its output, it can resist the plagiarism from other surrogate models to some extent. However, the biggest limitation of this method is that the added visible watermarks will sacrifice the visual quality and usability of the target model seriously. Another potential threat is that attackers may use image editing tools like Photoshops to manually remove all the visible watermarks.

\begin{figure}
    \centering
    \includegraphics[width=0.95\linewidth]{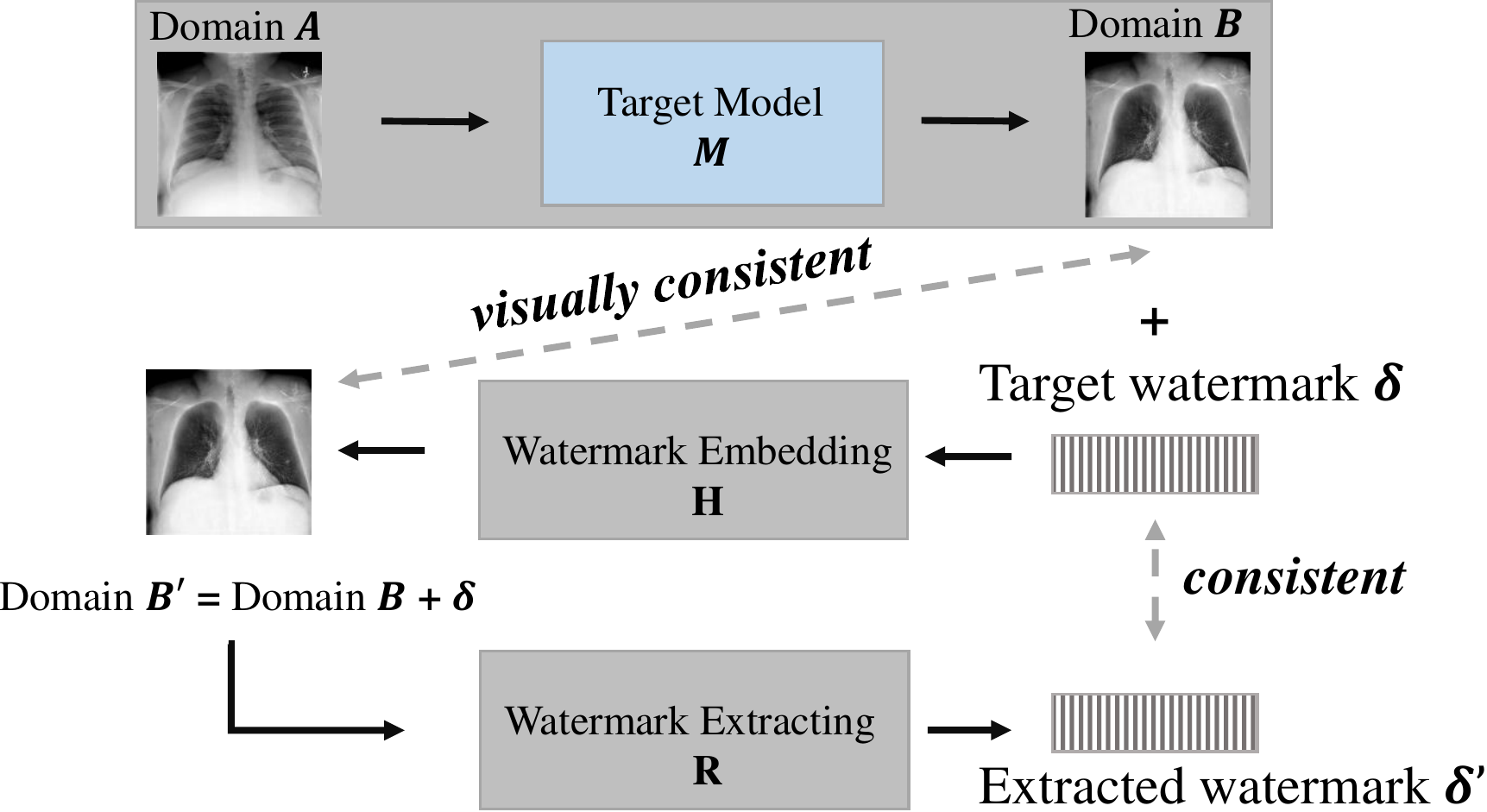}
    \caption{The proposed deep watermarking framework by leveraging spatial invisible watermarking algorithms.}
    \label{fig:framework}
\end{figure}

To address the above limitations, we propose a general model watermarking framework by leveraging the spatial invisible watermarking mechanism as shown in \Fref{fig:framework}. Given a target model $\mathbf{M}$ to be protected, we denote its original input and output images as domain $\mathbf{A}$  and $\mathbf{B}$ respectively. Then a spatial invisible watermark embedding method $\mathbf{H}$ is used to hide a unified target watermark $\bm{\delta}$ into all the output images in the domain $\mathbf{B}$ and generate a new domain $\mathbf{B'}$. Different from the above simple visible watermarks, all the images in the domain $\mathbf{B'}$ should be visually consistent to domain $\mathbf{B}$. Symmetrically, given the images in domain $\mathbf{B'}$, the corresponding watermark extracting algorithm $\mathbf{R}$ will extract the watermark $\bm{\delta'}$ out, which is consistent to $\bm{\delta}$. The key hypothesis here is that when the attacker uses $\mathbf{A}$ and $\mathbf{B'}$ to learn a surrogate model $\mathbf{SM}$, $\mathbf{R}$ can still extract the target watermark from the output $\mathbf{B''}$ of $\mathbf{SM}$.

We first test the effectiveness of our framework by using traditional spatial invisible watermarking algorithms like \cite{kutter1999watermarking,voloshynovskiy2000content}. It works well for some surrogate models but limited to some other ones. Another big limitation is that the information capacity they can hide is relatively low, e.g., tens of bits. To hide high capacity watermarks like logo images and achieve better robustness, we propose a novel deep invisible watermarking system shown in \Fref{fig:system_pipeline}, which consists of two main parts: one embedding sub-network $\mathbf{H}$ to learn how to hide invisible watermarks into the image, and another extractor sub-network $\mathbf{R}$ to learn how to extract the invisible watermark out. To avoid $\mathbf{R}$ generating watermark for all the images no matter whether they have invisible watermarks or not, we also constrain it not to extract any watermark out if its input is a clean image. To further boost the robustness, another adversarial training stage is used.

Experiments show that the proposed method can resist the attack from surrogate models trained with different network structures like Resnet and UNet and different loss functions like $L1$, $L2$, perceptual loss and adversarial loss. Depending on the specific task, we find it is also possible to combine the functionality of \textbf{M} and \textbf{H} to train a task-specific \textbf{H}.

To summarize, our contributions are fourfold:

\begin{itemize}
    \item We are the first to introduce the intellectual property protection problem for image processing tasks. We hope it can draw more attention to this research field and inspire greater works.
    \item We propose the first model watermarking framework to protect image processing networks by leveraging the spatial invisible watermarking mechanism.
    \item We design a novel deep watermarking algorithm to improve the robustness and capacity of traditional spatial invisible watermarking methods.
    \item Extensive experiments demonstrate that the proposed framework can resist the attack from surrogate models trained with different network structures and loss functions. It can also be easily extended to protect valuable data and traditional algorithms.
\end{itemize}

\section{Related work}
\noindent \textbf{Media Watermarking Algorithms.} Watermarking is one of the most important ways to protect media copyright. For image watermarking, many different algorithms have been proposed in the past decades, which can be roughly categorized into two types: visible watermarks like logos, and invisible watermarks. Compared to visible watermarks, invisible watermarks are more secure and robust. They are often embedded in the original spatial domain \cite{kutter1999watermarking,voloshynovskiy2000content,deguillaume2002method,voloshynovskiy2001multibit}, or other image transform domains such as discrete cosine transform (DCT) domain \cite{hsu1999hidden,hernandez2000dct}, discrete wavelet transform (DWT) domain \cite{barni2001improved}, and discrete Fourier transform (DFT) domain \cite{ruanaidh1996phase}. However, all these traditional watermarking algorithms are often only able to hide several or tens of bits, let alone explicit logo images. More importantly, we find only spatial domain watermarking work to some extent for this task and all other transform domain watermarking algorithms fail.

In recent years, some DNN-based watermarking schemes have been proposed. For example, Zhu \etal \cite{zhu2018hidden} propose an auto-encoder-based network architecture to realize the embedding and extracting process of watermarks. Based on it, Tancik \etal \cite{tancik2019stegastamp} further realize a camera shooting resilient watermarking scheme by adding a simulated camera shooting distortion to the noise layer. Compared to these image watermarking algorithms, model watermarking is much more challenging because of the exponential search space of deep models. But we innovatively find it possible to leverage spatial invisible watermarking techniques for model protection.

\noindent\textbf{Model Watermarking Algorithms.} Though watermarking for deep neural networks are still seriously under-studied, there are some recent works \cite{uchida2017embedding,adi2018turning,nagai2018digital,zhang2018protecting} that start paying attention to it. For example, based on the over-parameterized property of deep neural networks, Uchida \etal \cite{uchida2017embedding} propose a special weight regularizer to the objective function so that the distribution of weights can be resilient to attacks such as fine-tuning and pruning. One big limitation of this method is not task-agnostic and need to know the original network structure and parameters for retraining. Adi \etal \cite{adi2018turning} use a particular set of inputs as the indicators and let the model deliberately output specific incorrect labels, however, it may not work if the network are retrained. Zhang \etal \cite{zhang2018protecting} associate the watermark with the actual identity by making great changes to original images, which is easy to be detected. 

However, all the methods mentioned above focus on the classification tasks, which is different from the purpose of this paper: protecting higher commercial valued image processing models. We innovatively leverage spatial invisible watermarking algorithms for image processing networks and propose a new deep invisible watermarking technique to enable high-capacity watermarks(e.g., logo images).

\noindent\textbf{Image-to-image Translation Networks.}
In the deep learning era, most image processing tasks such as image segmentation, edge to the image, deraining, and X-ray image debone, can be solved with an image-to-image translation network where the input and output are both images. Recently this field has achieved significant progress especially after the emergence of the generative adversarial network (GAN) \cite{goodfellow2014generative}. Isola \etal propose a general image-to-image translation framework by combining adversarial training in \cite{pix2pix2017}, which is further improved by many following works \cite{choi2018stargan,wang2018high,park2019semantic}. The limitation of these methods is that they need a lot of pairwise training data. By introducing the cycle consistency, Zhu \etal propose a general unpaired image-to-image translation framework CycleGAN \cite{zhu2017unpaired}. In this paper, we mainly focus on the deep models of paired image-to-image translation, because the paired training data is much more expensive to be obtained than classification datasets or unpaired datasets. More importantly, there is no prior work that has ever considered the watermarking issue for such models.

\section{Method}
In this section, the details will be elaborated. Before that, we will first introduce the formal problem definition and give a simple theoretical pre-analysis to justify our hypothesis.

\noindent\textbf{Problem Definition.} For image processing tasks, assume the input domain $\mathbf{A}$ is composed of massive images $\{a_1, a_2, ..., a_n\}$, and the target output domain  $\mathbf{B}$ consists of images $\{b_1, b_2, ..., b_n\}$. In this paper, we only consider the pairwise case where $a_i$ and $b_i$ are one-one matched by an implicit transformation function $\mathcal{T}$. Then the goal of the image processing model $\mathbf{M}$ is to approximate $\mathcal{T}$ by minimizing the distance $\mathcal{L}$ between $\mathbf{M}(a_i)$ and $b_i$, i.e.,
\begin{equation}
\mathcal{L}(\mathbf{M}(a_i), b_i) \rightarrow 0.
\end{equation}

Assume we have learned a target model $\mathbf{M}$ based on massive private image pairs and computation resources. Given each input image $a_i$ in domain $\mathbf{A}$, $\mathbf{M}$ will output an image $b_i$ in domain $\mathbf{B}$. Then the attacker may use the image pairs defined by $(a_i, b_i)$ from domain $\mathbf{A},\mathbf{B}$ to train another surrogate model $\mathbf{SM}$. Our target is to design an effective deep watermarking mechanism which is able to identify $\mathbf{SM}$ once it is trained with data generated by $\mathbf{M}$. Because in real scenarios, it is highly possible we cannot access the raw model $\mathbf{SM}$ in a white-box way, the only indicator we can leverage is the output of $\mathbf{SM}$. Therefore, we need to figure out one way to extract watermarks from the output of $\mathbf{SM}$.

\noindent\textbf{Theoretical Pre-analysis.}
In traditional watermarking algorithms, given an image $I$ and a target watermark $\bm{\delta}$ to embed, they will first use a watermark embedding algorithm $\mathbf{H}$ to generate an image $I'$ which contains $\bm{\delta}$. Symmetrically, the target watermark $\bm{\delta}$ can be further extracted out with the corresponding watermarking extracting algorithm $\mathbf{R}$. Considering each image $b_i \in \mathbf{B}$ is embedded with a unified watermark $\bm{\delta}$, where $b'_i=b_i+\bm{\delta}$, forming another domain $\mathbf{B'}$, there must exist a model $\mathbf{M'}$ which can learn good transformation between domain $\mathbf{A}$ and $\mathbf{B'}$. One simplest solution of $\mathbf{M'}$ is to directly add $\bm{\delta}$ to the output of $\mathbf{M}$ with a skip connection:

\begin{equation}
\begin{aligned}
    \mathcal{L}(\mathbf{M}(a_i), b_i) & \rightarrow 0 \Leftrightarrow \mathcal{L}(\mathbf{M'}(a_i), (b_i+\mathbf{\delta})) \rightarrow 0 \\
   \mbox{when} \quad \mathbf{M'} &= \mathbf{M}(a_i) + \mathbf{\delta}.
\end{aligned}
\end{equation}

Based on the above observation, we propose a general deep watermarking framework for image processing models shown in \Fref{fig:framework}. Given a target model $\mathbf{M}$ to protect, we add a barrier $\mathbf{H}$ by embedding a unified watermark $\bm{\delta}$ into all its output images before showing them to the end-users. So the surrogate model $\mathbf{SM}$ has to be trained with the image pair $(a_i, b'_i)$ from domain $\mathbf{A},\mathbf{B'}$ with watermark, instead of the original pair $(a_i, b_i)$ from domain $\mathbf{A},\mathbf{B}$. No matter what architecture $\mathbf{SM}$ adopts, it's behavior will approach to that of $\mathbf{M'}$ in preserving the unified watermark $\bm{\delta}$. Otherwise, its objective loss function $\mathcal{L}$ cannot achieve a lower value. And then the watermarking extracting algorithm $\mathbf{R}$ can extract the watermark from the output of $\mathbf{SM}$.

To ensure the watermarked output image $b'_i$ is visually consistent with the original one $b_i$, only spatial invisible watermarking algorithms are considered in this paper. Below we will try both traditional spatial invisible watermarking algorithm and a novel deep invisible watermarking algorithm.

\noindent\textbf{Traditional Spatial Invisible Watermarking.} Additive-based embedding is the most common method used in the traditional spatial invisible watermarking scheme. The watermark is first spread to a sequence or block which satisfies a certain distribution, then embedded into the corresponding coefficients of the host image. This embedding procedure can be formulated by
\begin{equation}\label{TraSpatialMethod}
    I^{\prime}=\left\{\begin{array}{ll}{I+\alpha C_{0}} & {\text { if } w_i=0} \\ {I+\alpha C_{1}} & {\text { otherwise }}\end{array}\right.
\end{equation}
where $I$ and $I^{\prime}$ indicate the original image and embedded image respectively. $\alpha$ indicates the embedding intensity and $C_i$ denote the spread image block that represents bit ``$w_i$"($w_i\in[0,1]$). In the extraction side, the watermark is determined by detecting the distribution of the corresponding coefficients. The robustness of such an algorithm is guaranteed by the spread spectrum operation. The redundancy brought by the spread spectrum makes a strong error correction ability of the watermark so that the distribution of the block will not change a lot even after image processing. 

\begin{figure*}[!h]
\centering
\includegraphics[width=0.95\linewidth]{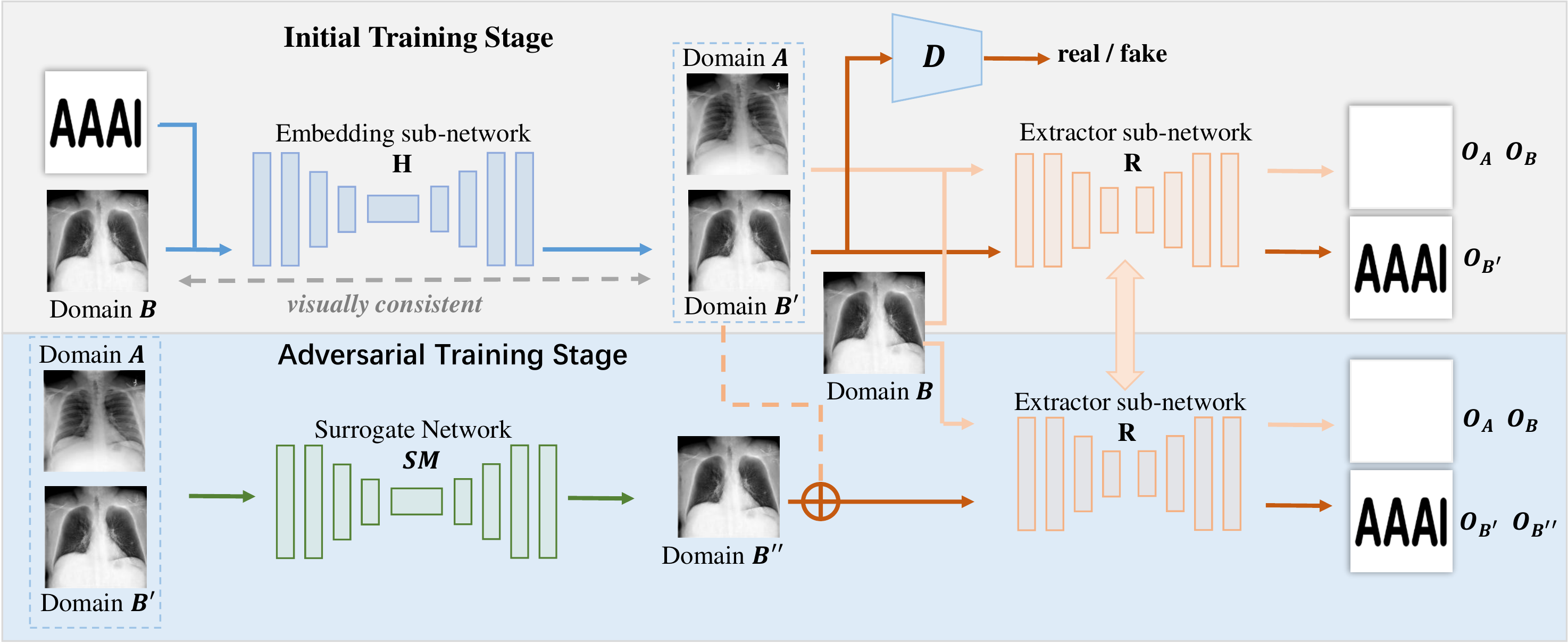}
\caption{The overall pipeline of the proposed deep invisible watermarking algorithm and two-stage training strategy. In the first training stage, a basic watermark embedding sub-network $\mathbf{H}$ and extractor sub-network $\mathbf{R}$ are trained. Then another surrogate network $\mathbf{SM}$ is leveraged as the adversarial competitor to further enhance the extracting ability of $\mathbf{R}$.}
\label{fig:system_pipeline}
\end{figure*}

However, such algorithms often have very limited embedding capacity because many extra redundant bits are needed to ensure robustness. In fact, in many application scenarios, the IP owners may want to embed some special images (e.g., logos) explicitly, which is nearly infeasible for these algorithms. More importantly, the following experiments show that these traditional algorithms can only resist some special types of surrogate models. To enable more high-capacity watermarks and more robust resistance ability, we propose a new deep invisible watermarking algorithm and utilize a two-stage training strategy as shown in \Fref{fig:system_pipeline}.

\noindent\textbf{Deep Invisible Watermarking.} To embed an image watermark into host images of the domain $\mathbf{B}$ and extract it out afterward, one embedding sub-network $\mathbf{H}$ and one extractor sub-network $\mathbf{R}$ are adopted respectively. Without sacrificing the original image quality of domain $\mathbf{B}$, we require images with the hidden watermark should be still visually consistent with the original images in the domain $\mathbf{B}$. Since adversarial networks have demonstrated their power in reducing the domain gap in many different tasks, we append one discriminator network $\mathbf{D}$ after $\mathbf{H}$ to further improve the image quality of domain $\mathbf{B'}$. During training, we find if the extractor network $\mathbf{R}$ is only trained with the images of domain $\mathbf{B'}$, it is very easy to overfit and output the target watermark no matter whether the input images contain watermarks or not. To avoid it, we also feed the images of domain $\mathbf{A}$ and domain $\mathbf{B}$ that do not contain watermarks into $\mathbf{R}$ and force it to output a constant blank image. In this way, $\mathbf{R}$ will have the real ability to extract watermarks only when the input image has the watermark in it.  

Based on the pre-analysis, when the attacker uses a surrogate model $\mathbf{SM}$ to imitate the target model $\mathbf{M}$ based on the input domain $\mathbf{A}$ and watermarked domain $\mathbf{B'}$, $\mathbf{SM}$ will learn the hidden watermark $\bm{\delta}$ into its output thanks to the inherent fitting property of deep networks. 
 Despite of higher hiding capacity, similar to traditional watermarking algorithms, the extractor sub-network $\mathbf{R}$ cannot extract the watermarks out from the output of the surrogate model $\mathbf{SM}$ neither if only with this initial training stage. This is because $\mathbf{R}$ has only observed clean watermarked images but not the watermarked images from surrogate models which may contain some unpleasant noises. To further enhance the extracting ability of $\mathbf{R}$, we choose one simple surrogate network to imitate the attackers' behavior and fine-tune $\mathbf{R}$ on the mixed dataset of domain $\mathbf{A}, \mathbf{B}, \mathbf{B'}, \mathbf{B''}$. Experiments show this will significantly boost the extracting ability of $\mathbf{R}$ and resist other types of surrogate models.

\noindent\textbf{Network Structures.}
In our method, we adopt the UNet \cite{ronneberger2015u} as the default network structure of $\mathbf{H}$ and $\mathbf{SM}$, which has been widely used by many translation based tasks like \cite{pix2pix2017,zhu2017unpaired}. It performs especially well for tasks where the output image shares some common properties of input image by multi-scale skip connections. But for the extractor sub-network $\mathbf{R}$ whose output is different from the input, we find CEILNet \cite{fan2017generic} works much better. It also follows an auto-encoder like network structure. In details, the encoder consists of three convolutional layers, and the decoder consists of one deconvolutional layer and two convolutional layers symmetrically. To enhance the learning capacity, nine residual blocks are inserted between the encoder and decoder. For the discriminator $\mathbf{D}$, we adopt the PatchGAN \cite{pix2pix2017} by default. Note that except for the extractor sub-network, we find other types of translation networks also work well in our framework, which demonstrates the strong generalization ability of our framework.

\noindent\textbf{Loss Functions.}
The objective loss function of our method consists of two parts: the embedding loss $\mathcal{L}_{emd}$ and the extracting loss $\mathcal{L}_{ext}$, i.e.,
\begin{equation}
    \mathcal{L} = \mathcal{L}_{emd} + \lambda*\mathcal{L}_{ext},
\end{equation}
where $\lambda$ is the hyper parameter to balance these two loss terms. Below we will introduce the detailed formulation of $\mathcal{L}_{emd}$ and $\mathcal{L}_{ext}$ respectively.

\noindent\textbf{Embedding Loss.}
To embed the watermark image while guaranteeing the original visual quality, three different types of visual consistency loss are considered: the basic $L$2 loss $\ell_{bs}$, perceptual loss $\ell_{vgg}$, and adversarial loss $\ell_{adv}$, i.e.,
\begin{equation}
\begin{aligned}
    \mathcal{L}_{emd} &= \lambda_1 * \ell_{bs} + \lambda_2 * \ell_{vgg} + \lambda_3 * \ell_{adv}.
\end{aligned}
\end{equation}
Here the basic $L$2 loss $\ell_{bs}$ is simply the pixel value difference between the input host image $b_i$  and  the watermarked output image $b_i'$, $N_c$ is the total pixel number, i.e.,
\begin{equation}
\begin{aligned}
    \ell_{bs} &= \sum_{b_i' \in \mathbf{B}', b_i \in \mathbf{B}} \frac{1}{N_c} \lVert b_i' - b_i\rVert^2. \\
\end{aligned}
\end{equation}
And the perceptual loss $\ell_{vgg}$ \cite{johnson2016perceptual} is defined as the difference between the VGG feature of $b_i$ and $b_i'$:
\begin{equation}
\begin{aligned}
    \ell_{vgg} &= \sum_{b_i' \in \mathbf{B}', b_i \in \mathbf{B}} \frac{1}{N_f} \lVert VGG_k(b_i') - VGG_k(b_i)\rVert^2,
\end{aligned}
\end{equation}
where $VGG_k(\cdot)$ denotes the features extracted at layer $k$ (``conv2\_2" by default), and $N_f$ denotes the total feature neuron number. To further improve the visual quality and minimize the domain gap between $\mathbf{B}'$ and $\mathbf{B}$, the adversarial loss $\ell_{adv}$ will let the embedding sub-network $\mathbf{H}$ hide watermarks better so that the discriminator $\mathbf{D}$ cannot differentiate its output from real watermark-free images in $\mathbf{B}$, i.e.,
\begin{equation}
\begin{aligned}
    \ell_{adv} &= \underset{b_i\in \mathbf{B}}{\mathbb{E}} log(\mathbf{D}(b_i)) + \underset{b_i'\in \mathbf{B'}}{\mathbb{E}} log(1 - \mathbf{D}(b_i')). \\
\end{aligned}
\end{equation}

\noindent\textbf{Extracting Loss.} The responsibility of the extractor sub-network $R$ has two aspects: it should be able to extract the target watermark out for watermarked images from $\mathbf{B'}, \mathbf{B''}$ and output a constant blank image for watermark-free images from $\mathbf{A,B}$. So the first two terms of $\mathcal{L}_{ext}$ are the reconstruction loss $\ell_{wm}$ and $\ell_{clean}$ for these two types of images respectively, i.e.,
\begin{equation}
\label{eq:wm}
    \begin{aligned}
    \ell_{wm} &= \sum_{b_i' \in \mathbf{B}'}\frac{1}{N_c}\lVert R(b_i') - \delta\rVert^2 + \sum_{b_i'' \in \mathbf{B}''}\frac{1}{N_c}\lVert R(b_i'') - \delta\rVert^2,  \\
    \ell_{clean} &= \sum_{a_i \in \mathbf{A}}\frac{1}{N_c}\lVert R(a_i) - \delta_0\rVert^2 + \sum_{b_i \in \mathbf{B}}\frac{1}{N_c}\lVert R(b_i) - \delta_0\rVert^2, \\
    \end{aligned}
\end{equation}
where $\delta_0$ is the constant blank watermark image. Besides reconstruction loss, we also want the watermarks extracted from different watermarked images to be consistent, thus another consistent loss is added:
\begin{equation}
\label{eq:cst}
    \ell_{cst} = \sum_{x,y \in \mathbf{B'}\cup\mathbf{B''}} \lVert R(x) - R(y)\rVert^2.
\end{equation}
Then $\mathcal{L}_{ext}$ is defined as the weighted sum of these three terms, i.e.,
\begin{equation}
    \mathcal{L}_{ext} = \lambda_4*\ell_{wm} + \lambda_5*\ell_{clean} + \lambda_6*\ell_{cst}.
\end{equation}

\noindent\textbf{Adversarial Training Stage.} With the above initial training stage, $\mathbf{R}$ only observes the clean watermarked images and cannot generalize well for the noisy watermarked output of some surrogate models. To enhance its extracting ability, an extra adversarial training stage is added. Specifically, one surrogate model $\mathbf{SM}$ is trained with the simple $L$2 loss by default. Denote the outputs of $\mathbf{SM}$ as $\mathbf{B}''$, we further fine-tune $\mathbf{R}$ on the mixed dataset $\mathbf{A, B, B', B''}$ in this stage. 

\section{Experiments}
In this paper, two examples of image processing tasks are conducted: image deraining and Chest X-ray image debone. The goal of these two tasks is to remove the rain streak and rib components from the input images respectively. To demonstrate the effectiveness of our method, we first show the newly introduced deep invisible watermarking algorithm can hide high-capacity image-based watermarks, then evaluate the robustness of the proposed deep watermarking framework to different surrogate models. Finally, some ablation analysis is provided to justify the motivation of our design and shed some light on more inspiring potentials.

\noindent\textbf{Implementation Details.} For image deraining, we use 12100 images from the PASCAL VOC dataset as target domain $\mathbf{B}$, and use the synthesis algorithm in \cite{zhang2018density} to generate rainy images as domain $\mathbf{A}$. These images are split into three parts: 
6000 both for the initial and adversarial training,  6000 to train the surrogate model
and 100 for testing. Similarly, for X-ray image debone, we select 6100 high-quality chest X-ray images from the open dataset chestx-ray8 \cite{wang2017chestx} and use the rib suppression algorithm proposed by \cite{yang2017cascade} to generate the training pair. They are also divided into three parts: 
3000 both for the initial and adversarial training, 3000 to train the surrogate model
and 100 for testing. By default, $\lambda,\lambda_1,\lambda_2,\lambda_4,\lambda_5,\lambda_6$ all equal to 1 and $\lambda_3=0.01$.

\noindent\textbf{Evaluation Metric.} To evaluate the visual quality, PSNR and SSIM are used by default. To judge whether the watermark is extracted successfully, we use the classic normalized correlation (NC) metric as previous watermarking methods. The watermark is regarded as successfully extracted if its NC value is bigger than $0.95$. Based on it, the success rate (SR) is further defined as the ratio of watermarked images whose hidden watermark is successfully extracted.

\begin{figure}
    \centering
    \includegraphics[width=0.9\linewidth]{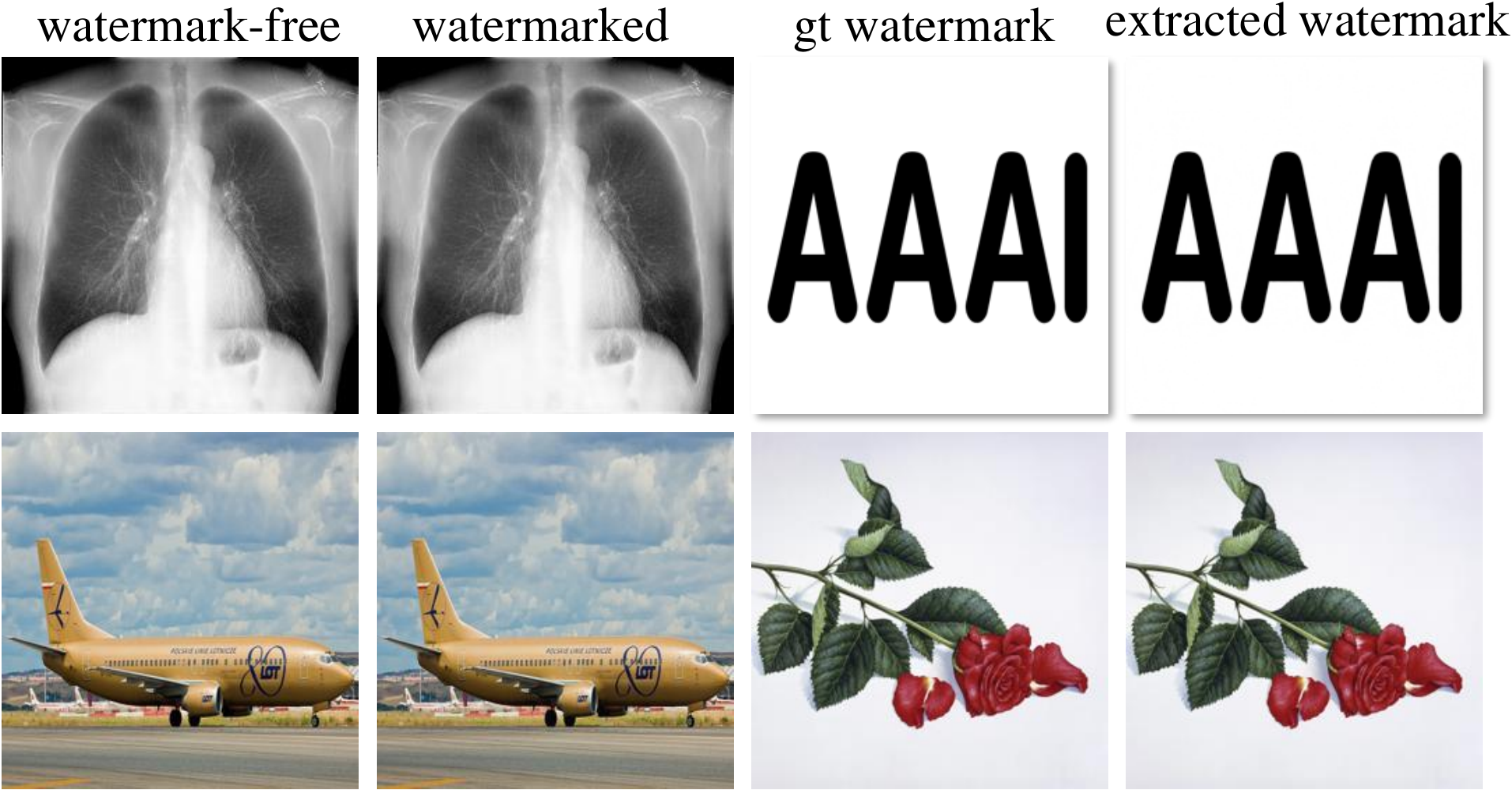}
    \caption{Two examples of hiding image watermark into host images with the proposed invisible watermarking algorithm.}
    \label{fig:invisible_wm_qual}
\end{figure}

\noindent\textbf{Deep Image Based Invisible Watermarking.} In this experiment, we give both quantitative and qualitative results about the proposed deep image based invisible watermarking algorithm. For debone and deraining, one AAAI logo image and a colorful flower image are used as the example watermark image respectively. As shown \Tref{tab:invisible_wm_quan}, the embedding sub-network $\mathbf{H}$ and the extractor sub-network $\mathbf{R}$ collaborate very well. $\mathbf{H}$ can hide the image watermark into the host images invisibly with high visual quality (average PSNR 39.98 and 47.89 for derain and debone respectively), and $\mathbf{R}$ can extract these hidden watermarks out afterward with an average NC value over 0.99 (100\% success rate). Two visual examples are shown in \Fref{fig:invisible_wm_qual}.

\begin{table}
    \centering
    \begin{tabular}{c|c|c|c}
    \hline
        Task & PSNR & SSIM & NC \\
        \hline
         Debone-aaai & 47.89 & 0.99 & 0.9999 \\
         \hline
         Derain-flower & 39.98 & 0.99 & 0.9966 \\
         \hline
    \end{tabular}
    \caption{Quantitative results of the proposed invisible image based watermarking. *-aaai and *-flower use the AAAI logo and a colorful flower image as watermarks respectively.}
    \label{tab:invisible_wm_quan}
\end{table}

\noindent \textbf{Robustness to The Attack from Surrogate Models.} To evaluate the final robustness of the proposed deep  watermarking framework, we use a lot of surrogate models that are trained with different network structures and objective loss functions to imitate the attackers' behavior. Here four different types of network structures are considered: vanilla convolutional networks only consisting of several convolutional layers (``CNet"), an auto-encoder like networks with 9 and 16 residual blocks (``Res9", ``Res16"), and the aforementioned UNet network (``UNet"). For objective loss functions, some popular loss functions like $L$1, $L$2, perceptual loss $L_{perc}$, adversarial loss $L_{adv}$ and their combination are considered. Since one surrogate model with ``UNet" and $L$2 loss function is leveraged in the adversarial training stage, this configuration can be viewed as a white-box attack and all other configurations are black-box attacks. 

\begin{table}
    \scriptsize
    \centering
    \setlength{\tabcolsep}{0.6mm}{
    \begin{tabular}{c|c|c|c|c|c|c}
        \hline
         Setting & T-Debone & T-Derain & D-Debone &
         D-Derain & D-Debone$\dagger$& D-Derain$\dagger$ \\
         \hline\hline
         CNet &0\%  &  0\%  &92\%  &100\%  &  0\%& 0\%\\
         \hline
         Res9 &0\% &    0\% &100\% &100\%  &  0\%& 0\%\\
         \hline
         Res16 &0\% &    0\%  & 100\% &100\%  & 0\% & 0\%\\
         \hline
         UNet &100\% &   100\%  &100\% &100\%  & 0\% & 0\%\\
         \hline

    \end{tabular}
    }
    \caption{The success rate (SR) of resisting the attack from surrogate models trained with $L$2 loss but different network structures. T-* means the results of using traditional spatial invisible watermarking algorithms to hide 64-bit, while D-* means that of the proposed deep invisible watermarking algorithm to hide watermark images. $\dagger$ denotes the results without adversarial training.}
    \label{tab:robustness_net}
\end{table}

Due to the limited computation resource, we do not consider all the combinations of different network structures and loss functions. Instead, we choose to conduct the control experiments to demonstrate the robustness to the network structures and loss functions respectively. In the \Tref{tab:robustness_net}, both traditional spatial bit-based invisible watermarking algorithms (hide 64-bit) and the proposed deep image-based invisible watermarking algorithm are tested. Though only UNet based surrogate model trained with $L2$ loss is leveraged in the adversarial training stage, we find the proposed deep model watermarking framework can resist both white-box and black-box attacks when equipped with the newly proposed deep image-based invisible watermarking technique. For traditional watermarking algorithms, they can only resist the attack of some special surrogate models because their extracting algorithms cannot handle noisy watermarked images from different surrogate models. More importantly, they cannot hide high-capacity watermarks like logo images. We have also tried many traditional \textit{transform domain watermarking algorithms} like DCT-based\cite{fang2018screen},DFT-based\cite{kang2010efficient} and {DWT-based\cite{kang2003dwt}} \textit{but all of them do not work and achieve $0\%$ success rate.}

To further demonstrate the robustness to different losses, we use the UNet as the default network structure and train surrogate models with different combinations of loss functions. As shown in \Tref{tab:robustness_loss}, the proposed deep watermarking framework has a very strong generalization ability and can resist different loss combinations with very high success rate. Since in real user scenarios, detailed network structure and training objective functions are the parts of the surrogate model that attacker may often change, we have enough reasons to think the proposed deep watermarking framework is applicable in these cases.

\begin{table}[t]
    \scriptsize
    \centering
    \setlength{\tabcolsep}{1mm}{
    \begin{tabular}{c|c|c|c|c|c|c}
        \hline
         Task & $L$1 & $L$1 + $L_{adv}$ & $L$2 & $L$2 + $L_{adv}$ & $L_{perc}$ & $L_{perc}$+$L_{adv}$ \\ \hline
         D-Debone & 100\% & 100\%  & 100\%  & 100\%  & 88\%  & 92\%  \\
         \hline
         D-Derain &  100\% &100\% & 100\% &100\%  &86\% &100\% \\
         \hline
         D-Debone$\dagger$ & 0\% & 98\%  & 0\%  & 100\%  & 0\%  & 0\%  \\
         \hline
         D-Derain$\dagger$ &  0\% & 0\% & 0\% &100\%  &24\% & 0\% \\
         \hline
    \end{tabular}
    }
    \caption{The success rate (SR) of resisting the attack from surrogate models that are trained with different loss combinations. $\dagger$ means the results without adversarial training.}
    \label{tab:robustness_loss}
\end{table}

\subsection{Ablation Study}

\begin{figure}[t]
    \centering
    \includegraphics[width=0.9\linewidth]{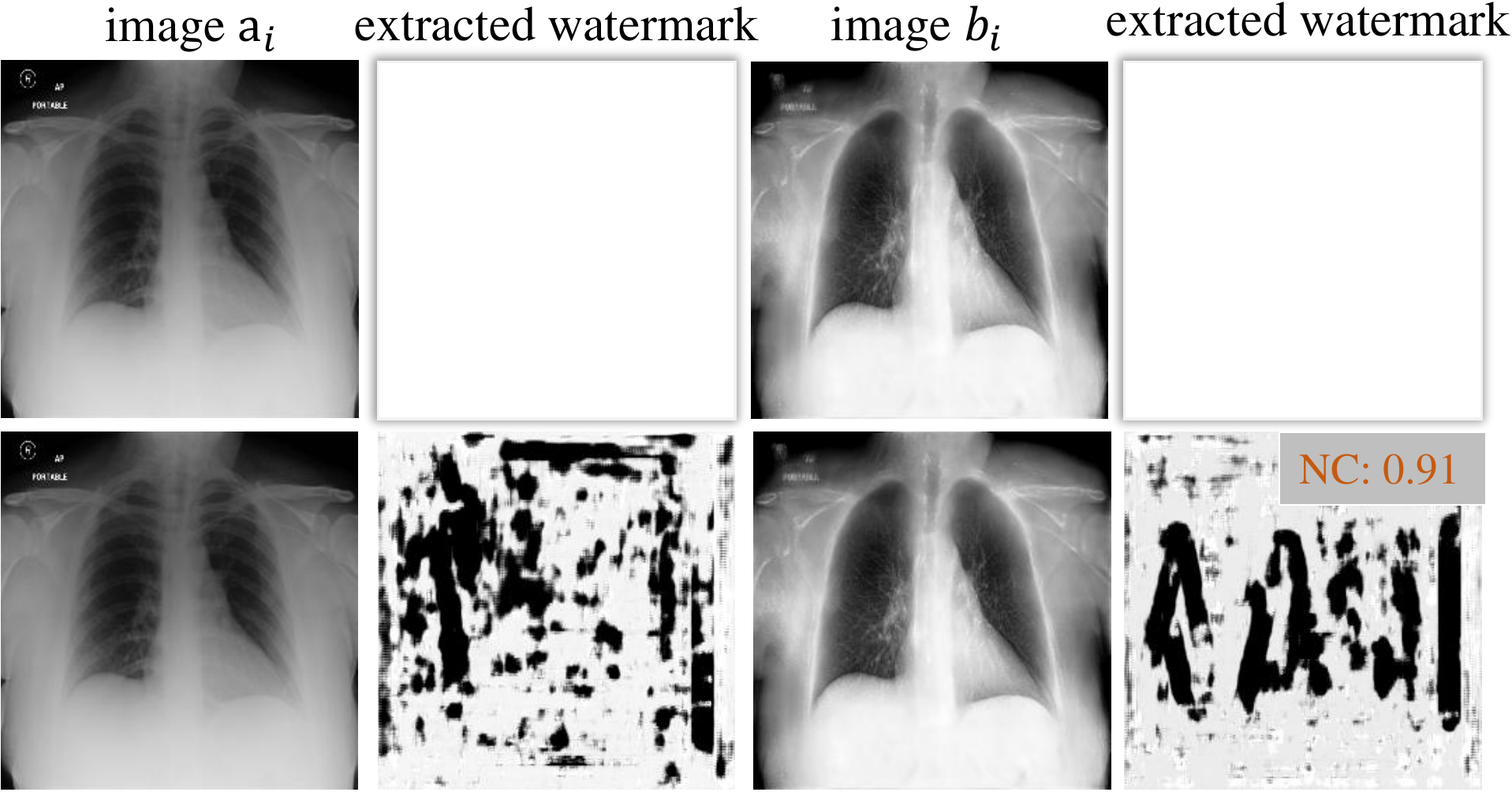}
    \caption{Comparison results with (first row) and without (second row) clean loss. The second and last column are the extracted watermarks from the watermark-free images $a_i, b_i$ from domain $\mathbf{A,B}$ respectively. }
    \label{fig:clean_loss}
\end{figure}

\begin{figure}[t]
    \centering
    \includegraphics[width=0.9\linewidth]{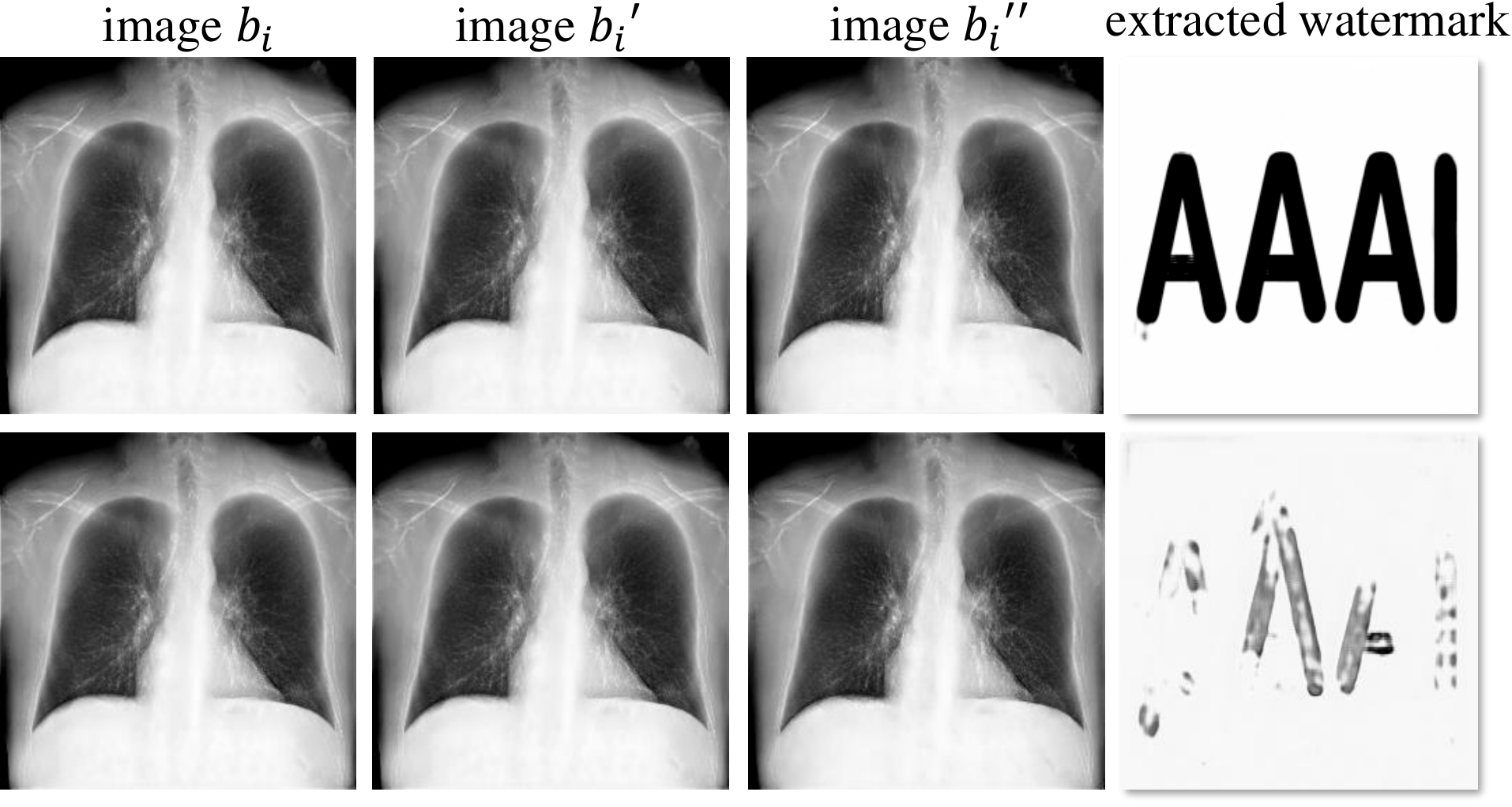}
    \caption{Comparison results with (first row) and without (second row) consistent loss. The last column is the extracted watermark from the output $b_i''$ of surrogate model. }
    \label{fig:consit_loss}
\end{figure}

\noindent \textbf{The Importance of Clean Loss and Consistent Loss.} Besides the watermark reconstruction loss, we add one clean loss and consistent loss into the extracting loss. To demonstrate their importance, two control experiments are conducted. As shown in \Fref{fig:clean_loss}, without clean loss, the extractor will always extract meaningless watermark from watermark-free images of domain $\mathbf{A,B}$. Especially for images of domain $\mathbf{B}$, the extracted watermarks have a quite large NC value and make the forensics meaningless. Similarly in \Fref{fig:consit_loss}, we find the extractor can only extract very weak watermarks or even cannot extract any watermark out when training without consistent loss. By contrast, our method can always extract very clear watermarks out.

\noindent \textbf{The Importance of Adversarial Training.} As described above, to enhance the extracting ability of $\mathbf{R}$, another adversarial training stage is used. To demonstrate its necessity, we also conduct the control experiments without adversarial training, and attach the corresponding results in  \Tref{tab:robustness_net} and \Tref{tab:robustness_loss} (labelled with $\dagger$). It can be seen that, with the default $L$2 loss, its resisting success rate is all about $0\%$ for surrogate models of different network structures. When using UNet as the network structure but training with different losses, we find only some special surrogate models can partially extract the hidden watermarks, which demonstrates the significant importance of the adversarial training. 

\noindent \textbf{Task Specific Deep Invisible Watermarks.} In our default setting, the embedding and extractor sub-network are task-agnostic and appended as a general barrier. In this experiment, we  try a more challenging task that lets $\mathbf{H}$ be task-specific. Take debone as the example, the input image of $\mathbf{H}$ is directly the image of the domain $\mathbf{A}$ with rib components now, and we want $\mathbf{H}$ to remove the rib components and hide watermarks simultaneously. In such a case, $\mathbf{H}$ is the final watermarked target model $\mathbf{M}$. For comparison, we also train a baseline debone model without the need of hiding watermarks. We find the above task-specific watermarked model can achieve very comparable results (PSNR: 24.49, SSIM: 0.91) to this baseline model (PSNR: 25.81, SSIM: 0.91).

\noindent \textbf{Extension to Protect Data and Traditional Algorithms.} Though our motivation is to protect deep models, the proposed framework is easy to be extended to protect valuable data or traditional algorithms by directly embedding the watermarks into their labeled groundtruth images or outputs.

\section{Conclusion}
We introduce the deep watermarking problem for image processing networks for the first time.
Inspired by traditional spatial invisible media watermarking, the first deep watermarking framework is proposed. To make it robust to different surrogate models and support image-based watermarks,  we 
propose a novel deep invisible watermarking technique. Experiments demonstrate that our framework can resist the attack from surrogate models trained with different network structures and loss functions. We hope this work can inspire more great works for this seriously under-researched field.

\subsection{Acknowledgments}
This work was partially supported by the Natural Science Foundation of
China under Grant U1636201, 61572452.

{
\bibliographystyle{aaai}
\bibliography{references}
}
\end{document}